\documentclass[12pt]{article}

\newcommand{\be}{\begin{equation}}
\newcommand{\ee}{\end{equation}}
\newcommand{\bea}{\begin{eqnarray}}
\newcommand{\eea}{\end{eqnarray}}

\newcommand{\gs}{\ensuremath{g_s}} 
\newcommand{\ap}{\ensuremath{\alpha'}} 
\newcommand{\ls}{\ensuremath{l_s}} 

\def\p{\partial}

\def\expec#1{\langle #1 \rangle}

\newcommand{\cF}{{\mathcal{F}}}

\newcommand{\cN}{{\mathcal{N}}}
\newcommand{\cO}{{\mathcal{O}}}
\newcommand{\cR}{{\mathcal{R}}}

\newcommand{\aD}{\ensuremath{\overline{\mbox{D3}}}}

\newcommand{\Nb}{\ensuremath{\bar{N}}}
\newcommand{\Tb}{\ensuremath{\bar{T}}}
\newcommand{\eb}{\ensuremath{\bar{e}}}
\newcommand{\Rb}{\ensuremath{\bar{R}}}
\newcommand{\rzb}{\ensuremath{\bar{r}_0}}
\newcommand{\mm}{\ensuremath{m_{\mathrm{FT}}}}
\newcommand{\mg}{\ensuremath{m_{\mathrm{SG}}}}
\newcommand{\sm}{\ensuremath{s_{\mathrm{FT}}}}
\newcommand{\Tm}{\ensuremath{T_{\mathrm{FT}}}}
\newcommand{\sg}{\ensuremath{s_{\mathrm{SG}}}}

\newcommand{\qm}{\ensuremath{Q_{\mathrm{FT}}}}
\newcommand{\qg}{\ensuremath{Q_{\mathrm{SG}}}}
\newcommand{\Pm}{\ensuremath{P^{(l)}_{\mathrm{FT}}}}
\newcommand{\Pg}{\ensuremath{P^{(l)}_{\mathrm{SG}}}}
\newcommand{\cm}{\ensuremath{\sigma^{(l)}_{\mathrm{FT}}}}
\newcommand{\cg}{\ensuremath{\sigma^{(l)}_{\mathrm{SG}}}}
\newcommand{\Gm}{\ensuremath{\Gamma^{(l)}_{\mathrm{FT}}}}
\newcommand{\Gg}{\ensuremath{\Gamma^{(l)}_{\mathrm{SG}}}}

\title{\bf Threebrane Absorption and Emission\\ from a
 Brane-Antibrane System}
\author{J.~Antonio Garc\'{\i}a\footnote{e-mail:
garcia@nuclecu.unam.mx}  ~and Alberto G\"uijosa\footnote{e-mail:
alberto@nuclecu.unam.mx}
\\{\small Departamento de F\'{\i}sica de Altas Energ\'{\i}as,
Instituto de Ciencias Nucleares}\\ {\small Universidad Nacional
Aut\'onoma de M\'exico}\\
{\small Apdo. Postal 70-543, M\'exico D.F. 04510}}
\date{}

\begin{document}
\maketitle
\begin{abstract}
We show that a previously proposed model based on a
D3-brane--anti-D3-brane system at finite temperature can reproduce
the low-frequency absorption and emission probabilities of the black
threebrane of Type IIB supergravity arbitrarily far from
extremality, for arbitrary partial waves of a minimal scalar
field.  Our calculations cover in particular the case of the neutral
threebrane, which corresponds to the Schwarzschild black hole in
seven dimensions.
Our results provide not only
significant evidence in favor of the
brane-antibrane model, but also a rationale for the
condition that the energies of the two component gases agree with one
another.
In the course of our analysis we correct
previous results on the absorption probabilities of the
near-extremal threebrane, and extend them to the far-from-extremal
regime.
\end{abstract}

\section{Introduction}

Recent work has opened the possibility to attain a quantitative
understanding of the physics of black holes \emph{far} from
extremality, an important and long-standing problem. In
particular, three years ago a study of brane-antibrane systems at
finite temperature led the authors of \cite{dgk,dgk2}
to construct a microscopic model for the black
threebrane of Type IIB supergravity and the black twobrane and fivebrane
of eleven-dimensional supergravity. The model is based on
decoupled stacks of branes and antibranes, with a gas of massless
particles on each stack, and was shown to successfully reproduce
the corresponding entropies arbitrarily far from
extremality.\footnote{For a list of other approaches to the
far-from-extremal problem, see, e.g., \cite{ghm}.} It also
correctly accounts for various other properties of the black
branes; in particular, their negative specific heat and pressure
find a natural explanation in terms of brane-antibrane
annihilation.

In the past few months, these results have been generalized in
various directions. It has been shown that the brane-antibrane
model predicts the correct entropy not only for the three
non-dilatonic cases studied in \cite{dgk}, but also for other
singly- and multiply-charged black branes
\cite{sp,bl,kalyanarama,lifschytz,ks,halyo} (see also the older
works \cite{hms,antientropy}). The model has been found to
reproduce even the highly non-trivial entropy formulas for branes
rotating with arbitrary amounts of angular momentum \cite{ghm,sp}.
Moreover, preliminary numerical calculations in the relevant
strongly-coupled gauge theory \cite{sp} (using the methods
developed in \cite{kll}) appear to support the result of
\cite{dgk} regarding the stability of the brane-antibrane system
at high enough temperatures, which is in turn one of the key
assumptions of the model.

In spite of this already quite significant body of evidence, there
are at least two aspects of the brane-antibrane model that are
still not fully understood. One is the fact that the model
accounts for the exact functional form of the entropy, but it
yields a numerical coefficient that,  at least under the
assumptions adopted in \cite{dgk}, is a power of two too small.
Intriguingly, in all (singly-charged) cases the discrepancy can be
summarized by the succinct but evidently unphysical statement that
the supergravity entropy behaves as if each of the gases in the
model had access to twice as much energy as is available to
it.\footnote{As shown in \cite{kalyanarama}, for multiply-charged
systems the energy carried by each gas would have to be a factor
$2^{1+(K-1)/\lambda}$ larger than is physically possible, where
$K$ denotes the number of charges and $\lambda$ specifies the
energy dependence of the gas entropy, $S\propto E^{\lambda}$.}
What is needed then is a physically plausible reinterpretation of
this seemingly simple pattern.

In \cite{sp} it was noted that it is best to trace the discrepancy
back to a disagreement between the supergravity and microscopic
\emph{masses} (rather than entropies), which would in turn imply
that some type of binding energy has not been properly taken into
account in the microscopic side. This point of view is physically
sensible, and we will have more to say about it and a closely
related possibility in the Conclusions. Nevertheless, given the
simple form of the discrepancy one would hope to be able to pin
down its origin quantitatively.

In \cite{kalyanarama,ks} a different, `empirical' interpretation
of the discrepancy was proposed, in terms of the allocation of the
total gas energy to only \emph{one} of the gases (or the
corresponding fraction of \emph{each}), and an accompanying
reduction by a factor of four in the value of the component brane
tension. Again, to be satisfied one would like to derive both of
these conditions dynamically. For instance, one could try to
associate the reduction in the tension with \emph{partial}
brane-antibrane annihilation, which could then also conceivably be
responsible for the reduction in the number of gas degrees of
freedom. But such an interpretation would run into trouble for the
charged black brane, because in that case the number of branes and
antibranes is not equal, and the surplus branes (or antibranes)
would not be able to annihilate. Moreover, as already mentioned,
the results of \cite{dgk,sp} appear to indicate that the
brane-antibrane system is in fact stable, and therefore no such
partial annihilation should take place. It seems then that one
must regard the proposed reduction in brane tension as originating
from binding energy, which essentially brings us back to the
interpretation of \cite{sp}, except that here we are in addition
missing an explanation for the putative `deactivation' of all but
one of the gases.

A second (and possibly more significant) puzzling feature of the
model is the fact that, to correctly reproduce the entropy of
black branes with non-zero charge, the gases involved must be
assumed to have equal energies, and consequently different
temperatures. Since the gases are assumed to be decoupled from one
another, their having different temperatures is not entirely out
of the question, but up to now we do not understand why it is that
their energies (or equivalently, their pressures) must
agree.\footnote{In the interpretation of \cite{kalyanarama,ks},
the condition that the gas energies be the same is replaced by the
equally puzzling condition that only one of the gases
contributes.} In other words, unless there is some physical
restriction, in the microscopic side one could construct a family
of systems which differ only in the way the total gas energy is
split among the component gases, and each of these systems would
be expected to have a supergravity counterpart. Families of
\emph{regular} supergravity solutions with the desired properties
are not known to exist, so in \cite{dgk} it was speculated that
perhaps the condition that the gas energies be equal in the
microscopic side could somehow correspond to the absence of
singularities in the supergravity side.\footnote{This possibility
has been raised independently by David Lowe.}

Clearly the brane-antibrane model will not be on firm ground until
these two aspects are properly understood. In the meantime, and as
part of the strategy to achieve that goal, it is important to
subject the model to further tests, and in particular to compute
quantities other than the entropy.  Initial steps in this
direction were taken already in \cite{dgk}, where it was shown
that the model implies the correct form for the supergravity
energy-momentum tensor, and a transverse size (due to thermal
fluctuations) for the microscopic system which is of the same
order as the horizon radius. (This last property has also been
seen in \cite{sp,bl,halyo}, and is in consonance with the proposal of
\cite{mathur}.) In that same work, a study of the thermodynamic
counterpart \cite{glthermo} of the Gregory-Laflamme instability
\cite{gl} led to the satisfying conclusion that the microscopic
system, while unstable to collapse when wrapped on a sufficiently
large torus, would actually stabilize at a finite size and acquire
properties coinciding with those of a ten- or eleven-dimensional
black \emph{hole}. Very recently, the thermalization rate of the
brane-antibrane system was studied in \cite{lifschytz2}, and
compared against the quasi-particle picture developed in
\cite{ikll} (see also \cite{halyobits}), again with satisfying
results.

In this paper we will take additional steps in this direction, by
studying the manner in which the microscopic system absorbs and
emits radiation. More specifically, we will compare the absorption
probabilities and Hawking radiation rates predicted by the
brane-antibrane model against the actual supergravity results, to
lowest order in the radiation frequency. For concreteness, we will
restrict our analysis to radiation associated with a minimal
scalar field, in the presence of a black threebrane with arbitrary
charge. Given the successful generalizations in
\cite{ghm,sp,bl,kalyanarama,lifschytz,ks,halyo}, we would expect
analogous results for other types of black brane.

We start in Section \ref{reviewsec} by reviewing the results of
the brane-antibrane model for the threebrane case. We next work
out in Section \ref{microsec} the absorption probabilities for the
microscopic system. This requires an analysis of the corresponding
probabilities in the case of a \emph{near-extremal} threebrane,
which are determined in Section \ref{symsubsec}, correcting
previous results. The absorption probabilities predicted by the
model are then explicitly written down in Section
\ref{predictsubsec}, and shown to agree with their supergravity
counterparts in Section \ref{sugrasec}, first for the previously
examined neutral case in \S \ref{neutralsubsec}, and then for the
newly computed charged case in \S \ref{chargedsubsec}. Finally, in
Section \ref{emissionsec} we carry out a successful comparison
between the rates of Hawking radiation in the microscopic and
supergravity sides. We conclude in Section \ref{conclsec}, which
includes both a summary of our results in \S \ref{summarysubsec}
and a critical discussion on the assumptions of the model in \S
\ref{criticalsubsec}.

\section{Review of the Brane-Antibrane Model} \label{reviewsec}

The metric of the black threebrane solution of Type IIB
supergravity takes the form
$$
ds^2={1\over\sqrt{H(r)}}\left(-f(r)dt^2+d\vec{x}\,^2\right)
+\sqrt{H(r)}\left({dr^2\over f(r)}+ r^2 d\Omega_5^2\right),$$
where
$$
H(r)=1+\frac{r_h^4\sinh^2\alpha}{r^4}~,\qquad f(r)=1-{r_h^4\over
r^4}~,
$$
with $r_{h}$ the horizon radius. The ADM mass density of this
geometry is
\be \label{mg}
\mg\equiv {M_{\mathrm{SG}}\over
V}={\pi^3\over\kappa^2}r_h^4\left({3\over 2}+\cosh 2\alpha
\right),
\ee
its entropy density
\be \label{sg}
\sg\equiv {A_{h}/4G_{N}\over
V}={2\pi^2\over\kappa^2}r_h^5\cosh\alpha~,
\ee
and its Hawking temperature
\be \label{tg}
T_H={1\over\pi r_h\cosh\alpha}~.
\ee
The solution also involves a RR five-form field-strength,
associated with a charge
\be \label{qg}
\qg={\pi^{5/2}\over\kappa}r_h^4\sinh 2\alpha~.
\ee

It was shown in \cite{dgk} that $\sg(\mg,\qg)$ can be reproduced
with a field-theoretic model based on a system of $N$ D3-branes,
$\Nb$ anti-D3-branes, and two gases of $\cN=4$ super-Yang-Mills
(SYM) particles, arising respectively from the massless modes of
3-3 and $\overline{3}$-$\overline{3}$ open strings. The stack of
branes and that of antibranes (together with their corresponding
gases) are assumed not to interact with one another. The charge of
the microscopic system is then
\be \label{qm}
\qm=N-\Nb~,
\ee
and its total mass density,
\be \label{mm}
\mm=(N+\Nb)\tau_{3}+e+\eb~,
\ee
with $\tau_{3}=\sqrt{\pi}/\kappa$ the D3-brane tension and $e$
($\eb$) the energy density of the gas on the D3-branes
(\aD-branes). As mentioned in the Introduction, the model assumes
that $e=\eb$.

The entropy of the system is entirely due to the two $\cN=4$ SYM
gases. In the regime of interest, $\gs (N-\Nb)\gg 1$ (where the
supergravity solution is reliable), SYM is strongly-coupled, and
so the entropy of the two gases cannot be determined
perturbatively. It is however known \cite{gkp,threequarters} via
the AdS/CFT correspondence \cite{malda},
\be \label{sm}
\sm=2^{5/4}3^{-3/4}\pi^{1/2}\left(e^{3/4}\sqrt{N}+\eb^{3/4}\sqrt{\Nb}
\right).
\ee
Equations (\ref{qm}) and (\ref{mm}) can  be used in (\ref{sm})
to eliminate $\Nb$ and $e$ in favor of $N$, and the optimal value
of $N$ determined by maximizing $\sm$ at fixed $\qm$ and $\mm$.
Requiring that $\qm=\qg$ and $\mm=\mg$, the resulting equilibrium
values of $N$, $\Nb$ and $e$ can then be expressed in terms of the
supergravity parameters $r_h$ and $\alpha$:
\be \label{nnbar}
N={\pi^{5/2}\over 2\kappa}r_h^4 e^{2\alpha}, \qquad
\Nb={\pi^{5/2}\over 2\kappa}r_h^4 e^{-2\alpha}, \qquad
e=\eb={3\pi^3\over 4\kappa^2}r_h^4~.
\ee
Inserting these expressions into (\ref{sm}), we obtain agreement
with the supergravity entropy (\ref{sg}), up to a numerical
coefficient: $\sg=2^{3/4}\sm$. Given the dependence (\ref{sm}) of
the microscopic entropy on the gas energy, we see that, as noted
in the Introduction, the supergravity entropy behaves as if
\emph{each} gas carried \emph{twice} the available energy.

For later use, let us also recall that the energies of the two SYM
gases are related to the corresponding temperatures through
\be \label{em}
e={3\pi^2\over 8}N^2 T^4, \qquad \eb={3\pi^2\over 8}\Nb^2 \Tb^4~,
\ee
so at equilibrium we have
\be \label{tm}
T={2^{3/4}\over\pi r_h e^{\alpha}} , \qquad \Tb={2^{3/4}\over\pi
r_h e^{-\alpha}} ~.
\ee
As noted in \cite{dgk}, the overall temperature of the system,
$\Tm\equiv(\p\sm/\p\mm)^{-1}_{\qm}$, can be expressed in
terms of the gas
temperatures through
$$
{2\over\Tm}={1\over T}+{1\over\Tb}~,
$$
and is therefore a factor of $2^{3/4}$ larger than the Hawking
temperature (\ref{tg}), as expected from the numerical discrepancy
between $\sm$ and $\sg$.

\section{Microscopic Absorption Probabilities}
\label{microsec}

In the model the stack of branes is decoupled from the stack of
antibranes, so, at least for low enough frequencies,
the probability that the system absorbs quanta of a
given field with frequency $\omega$ must be given by the sum of
the two independent contributions,
\be \label{pmsum}
\Pm(\omega)=P^{(l)}(\omega;N,T)+P^{(l)}(\omega;\Nb,\Tb),
\ee
where $P^{(l)}(\omega;N,T)$ denotes the probability of absorption by a
strongly-coupled $SU(N)$ SYM gas with temperature $T$. In the next
subsection we will determine this probability. For simplicity, we
will consider only absorption of a minimal scalar field with low
frequency, in the sense that
\be \label{lowfreq}
\omega\ll T,\Tb~,
\ee
which through (\ref{tm}) is seen to imply $\omega\ll 1/r_{h}$.

\subsection{Absorption by a strongly-coupled $\cN=4$ SYM gas\\
(or, equivalently, by a near-extremal threebrane)}
\label{symsubsec}

Just as was done for the entropy calculation in \cite{dgk}, we
will use the AdS/CFT correspondence \cite{malda} to map the SYM
absorption calculation onto the problem of absorption by a
\emph{near-extremal} black threebrane. The latter's throat radius
$R$ and horizon radius $r_0$ are related to the gauge group rank
$N$ and gas temperature $T$ through the well-known expressions
\be \label{Rr0}
R^4=4\pi\gs N\ls^4={\kappa N\over 2\pi^{5/2}}~, \qquad r_0=\pi T
R^2=T\sqrt{\kappa N\over 2\pi^{1/2}}~.
\ee
Of course, these formulas are simply a rewriting of the general
expressions (\ref{qg}) and (\ref{tg}) in the near-extremal limit,
$r_h\to 0$ with $r_h^2 e^{\alpha}$ fixed. \emph{It is very
important, however, not to confuse $r_0$ and $R^4$, which should
be regarded as auxiliary parameters in the microscopic
calculation, with $r_h$ and $r_h^4\sinh^2\alpha$, which are the
parameters that characterize the arbitrarily far-from-extremal
black brane whose absorption probabilities we will attempt to
reproduce.}

The absorption probability for a minimal scalar field $\phi$ with
frequency (\ref{lowfreq}) on the background of a near-extremal
black threebrane has been computed in \cite{ps}.\footnote{The same
work also computes the probability in the opposite regime,
$\omega\gg T$, correcting the previous results \cite{d3absorption}.}
There are a few errors and misprints in that calculation, however,
that we will now correct.

The radial equation of motion for the $l$th partial wave is
\be \label{rhoeq}
\p^2_{\rho}\phi+\frac{5\rho^4-\rho_0^4}{\rho(\rho^4-\rho_0^4)}\p_{\rho}\phi
-\frac{\rho^2 l(l+4)}{\rho^4-\rho_0^4}\phi +
\frac{\rho^4(\rho^4+\cR^4)}{(\rho^4-\rho_0^4)^2}\phi=0~,
\ee
where following \cite{ps} we have defined
\be \label{rhos}
\rho=\omega r~,\qquad \rho_0=\omega r_0~, \qquad \cR=\omega R~.
\ee
The calculation in \cite{ps} assumes that
\be \label{regime}
\rho_0\ll\cR\ll 1~.
\ee
In the \emph{outer region} $\rho\gg\rho_0$, (\ref{rhoeq}) reduces
to
\be \label{rhoeqouter}
\p^2_{\rho}\phi+\frac{5}{\rho}\p_{\rho}\phi +
\left(1+{\cR^4\over\rho^4}-\frac{ l(l+4)}{\rho^2}\right)\phi=0~.
\ee
Notice that the ratio between the second and third terms inside
the large parentheses is of order
$$
{\cR^4\over\rho^2}=\left({\omega R^2\over r}\right)^2\ll
\left({\omega R^2\over r_0}\right)^2=\left({\omega \over \pi
T}\right)^2\ll 1~,
$$
where we have made use of (\ref{Rr0}) and (\ref{lowfreq}). So in
the outer region, \emph{independently} of whether or not $r\gg R$
($\rho\gg\cR$), the $\cR^4$ term in (\ref{rhoeqouter}) can be
dropped.\footnote{In \cite{ps}, the outer region is defined
instead as $r\gg R$, but then it would not overlap with the inner
region, which, as we will see below (and contrary to what is
stated in \cite{ps}), must be restricted to $r\ll R$.} The first
term inside the large parentheses, on the other hand, will be
relevant for $\rho\gg 1$, and so cannot be dropped. The resulting
equation is related to Bessel's equation, and the general solution
can be written as
\be \label{besselsol}
\phi(\rho)={A_l\over\rho^2}J_{l+2}(\rho)+{B_l\over\rho^2}N_{l+2}(\rho)~.
\ee
Using the asymptotic form of the Bessel functions for $\rho\gg 1$,
the ingoing flux at infinity then follows as
\bea \label{farflux}
\cF^{(\mathrm{in})}_{r\to\infty}&\equiv& \lim_{r\to\infty}
{f(r)r^5\over
2i}\left(\phi^{(\mathrm{in})*}\p_r\phi^{(\mathrm{in})}
-\phi^{(\mathrm{in})}\p_r\phi^{(\mathrm{in})*}\right)\nonumber\\
{}&=&-\frac{|A_l+iB_l|^2}{2\pi\omega^4}~.
\eea
On the other hand, for $\rho\ll 1$ (which includes in particular
the region $\rho_0\ll\rho\ll\cR$), we have
\be \label{innerbessel}
\phi(\rho)={A_l\rho^l\over
2^{l+2}(l+2)!}-{B_l\,2^{l+2}(l+1)!\over\pi\rho^{l+1}}~.
\ee

Consider now the \emph{inner region} $\rho_0\le\rho\ll \cR\ll 1$.
Defining $x=\rho_0^2/\rho^2$ as in \cite{ps}, (\ref{rhoeq}) can be
rewritten as
\be \label{rhoeq2}
\p_x^2\phi-\frac{1+x^2}{x(1-x^2)}\p_x\phi-
\frac{l(l+4)}{4x^2(1-x^2)}\phi+\frac{\cR^4/\rho_0^2
+\rho_0^2/x^2}{4x(1-x^2)^2}\phi=0~.
\ee
The $\rho_0^2/x^2$ term can be neglected in comparison with
$\cR^4/\rho_0^2$, since this just amounts to the statement that
$\rho\ll\cR$, which defines the inner region.\footnote{As mentioned in
the previous footnote, in \cite{ps} the inner region is defined
simply as $\rho_{0}\le\rho\ll 1$, and it is incorrectly stated that throughout
this region
$\rho_0^2/x^2$ can be dropped compared to $\cR^4/\rho_0^2$.}
At the same time, by
assumption we know that $\cR^4/\rho_0^2=(\omega/\pi T)^2\ll 1$,
and so the last term in (\ref{rhoeq2}) is seen to be completely
irrelevant unless one is very close to the horizon, $x\simeq 1$,
where it gives the dominant contribution and implies that
$\phi(\rho)\propto(1-x^2)^{\pm i\cR^4/4\rho_0^2}$. Choosing the
lower sign in order for the solution to be purely ingoing at the
horizon, one is thus led to the conclusion that
\be \label{phiansatz}
\phi(\rho)=(1-x^2)^{- i\cR^2/4\rho_0}\varphi(x)~,
\ee
where, to leading order in $\cR^2/\rho_0$, $\varphi(x)$ satisfies
equation (\ref{rhoeq2}) with the last term omitted, and is by
construction regular at the horizon, $x=1$.

Given (\ref{phiansatz}), the ingoing flux at the horizon follows
as
\bea \label{horflux}
\cF^{(\mathrm{in})}_{r\to r_{0}}&\equiv&\lim_{r\to r_{0}}
{f(r)r^5\over 2i}\left(\phi^{*}\p_r\phi-\phi\p_r\phi^{*}\right)\nonumber\\
{}&=&-\frac{\cR^{2}\rho_{0}^{3}|\varphi(1)|^2}{\omega^4}~.
\eea
Combining (\ref{farflux}) and (\ref{horflux}) we can write down
the desired absorption probability\footnote{The factor
$|\varphi(1)|^{2}$ was erroneously omitted from the calculation in
\cite{ps}, so their absorption probabilities are off by this
($l$-dependent) constant.}
\be \label{pabs}
P^{(l)}(\omega;N,T)\equiv\frac{\cF^{(\mathrm{in})}_{r\to r_{0}}}
{\cF^{(\mathrm{in})}_{r\to\infty}}=2\pi\cR^{2}\rho_{0}^{3}
\frac{|\varphi(1)|^{2}}{|A_l+iB_l|^{2}}~.
\ee

To complete the calculation, we need to determine $\varphi(1)$ and
$\varphi(x\ll 1)$ (which compared against (\ref{innerbessel}) will
allow us to identify $A_l$ and $B_l$). As explained in \cite{ps},
the function $\varphi(x)$ can be expressed in terms of
hypergeometric functions:
\bea \label{hypersol}
\varphi(x)&=& D_{l}\varphi_1(x)+C_{l}\varphi_2(x)~,\\
\varphi_{1}(x)&=&x^{2+l/2}F(1+l/4,1+l/4;2+l/2;x^{2})~,\nonumber\\
\varphi_{2}(x)&=&x^{-l/2}F(-l/4,-l/4;-l/2;x^{2})~.
\eea
The coefficients $D_{l},C_{l}$ must be determined by requiring
$\varphi$ to be smooth at $x=1$, with help of the relation
\bea \label{hyperx1}
F(a,b;c;z)&=&\frac{\Gamma(c)\Gamma(c-a-b)}
{\Gamma(c-a)\Gamma(c-b)}F(a,b;a+b
-c+1;1-z) \nonumber\\
{}&{}&+(1-z)^{c-a-b}
\frac{\Gamma(c)\Gamma(a+b-c)}
{\Gamma(a)\Gamma(b)}F(c-a,c-b;c-a-b
+1;1-z). \nonumber
\eea
Since in our case $c=a+b$, the above relation must be regularized by
letting $c\to c+\epsilon$. Depending on the value of $l$, there are
three different cases to consider \cite{ps}:
\begin{itemize}
 \item {\bf For odd values of $l$}, the
  $\log(1-x^{2})$ singularities in $\varphi_{1}(x\to 1)$ and
  $\varphi_{2}(x\to 1)$ cancel if we choose
  $$
  D_{l}=1, \qquad C_{l}=-\frac{\Gamma(2+l/2)\Gamma(-l/4)^{2}}
    {\Gamma(1+l/4)^{2}\Gamma(-l/2)}~.
  $$
  One can then deduce that
  $$
  \varphi(1)=(-1)^{(l-1)/2}2\pi\frac{\Gamma(2+l/2)}
  {\Gamma(1+l/4)^{2}}~,
  $$
  whereas for $x\ll 1$ ($\rho\gg\rho_{0}$) we have
  $$
  \phi(x)\simeq\varphi(x)\simeq C_{l}(\rho/\rho_{0})^{l}~.
  $$
  Matching this with the outer region's solution (\ref{innerbessel}),
  we see that to this order
  $$
  A_l\simeq 2^{l+2}(l+2)!C_{l}\rho_{0}^{-l}, \qquad B_l\simeq 0~.
  $$
  Employing this and the value of $\phi(1)$ in the master formula
  (\ref{pabs}), we finally conclude that
  \be \label{podd}
  P^{(l)}(\omega;N,T)={2^{-2l-1}\pi^{3}\over(l+2)!^{2}}\frac{\Gamma(-l/2)^{2}}
  {\Gamma(-l/4)^{4}}\omega^{2l+5}r_{0}^{2l+3}R^{2}~,
  \ee
  with $r_{0}(N,T)$ and $R(N)$ given by (\ref{Rr0}).
 \item {\bf For $l$ and $l/2$ even}, $\varphi_{1}$ still has a
  logarithmic singularity, but
  $\varphi_{2}$ becomes a Legendre polynomial,
  $$
  \varphi_{2}(x)={n!^{2}\over(2n)!}P_{n}(2/x^{2}-1)~.
  $$
  We can then set $D_{l}=0$, $C_{l}=1$, and deduce that
  $$
  \varphi(1)={n!^{2}\over(2n)!}
  $$
  and
  $$
  \varphi(x\ll 1)\simeq (\rho/\rho_{0})^{l}\quad
  \Longrightarrow\qquad A_l\simeq 2^{l+2}(l+2)!\rho_{0}^{-l},
  \quad B_l\simeq 0~.
  $$
  So in this case
  \be \label{peveneven}
  P^{(l)}(\omega;N,T)={2^{-2l-3}\pi\over(l+2)!^{2}}\frac{(l/4)!^{4}}{(l/2)!^{2}}
  \omega^{2l+5}r_{0}^{2l+3}R^{2}~.
  \ee
 \item {\bf For $l$ even, $l/2$ odd}, $\varphi_{1}$ is still singular
  and $\varphi_{2}$ is ill-defined (and its regularized version is
  proportional to $\varphi_{1}$). The desired non-singular solution can be
  expressed in terms of Meijer's $G$ function \cite{gr},
  $$
  \varphi(x)=G_{22}^{20}\left(x^{2}\left|{1 \atop -l/4}\quad{1\atop
  1+l/4}\right.\right)~,
  $$
  which implies that
  $$
  \varphi(1)=1
  $$
  and\footnote{The $(l/2)!$ factor is missing in \cite{ps}.}
  $$
  \varphi(x\ll 1)\simeq \frac{(l/2)!}{\Gamma(1+l/4)^{2}}
  (\rho/\rho_{0})^{l}\quad
  \Longrightarrow\qquad A_l\simeq \frac{2^{l+2}(l+2)!(l/2)!}
  {\Gamma(1+l/4)^{2}}\rho_{0}^{-l}, \quad B_l\simeq 0~.
  $$
  It follows that
  \be \label{pevenodd}
  P^{(l)}(\omega;N,T)={2^{-2l-3}\pi\over(l+2)!^{2}}\frac{\Gamma(1+l/4)^{4}}
  {(l/2)!^{2}}\omega^{2l+5}r_{0}^{2l+3}R^{2}~,
  \ee
  which is in fact the same formula as (\ref{peveneven}).
\end{itemize}

Using the identity $\Gamma(x)=\pi/[\sin(\pi x)\Gamma(1-x)]$, the
result (\ref{podd}) for odd $l$ can be put in the form
\be \label{pmgral}
P^{(l)}(\omega;N,T)={2^{-2l-3}\pi\over(l+2)!^{2}}\frac{\Gamma(1+l/4)^{4}}
{\Gamma(1+l/2)^{2}}\omega^{2l+5}r_{0}^{2l+3}R^{2}~,
\ee
which agrees with (\ref{pevenodd}) and is therefore seen to hold
for all values of $l$.

\subsection{Predictions of the brane-antibrane model}
\label{predictsubsec}

In the previous subsection we have seen that the absorption
probability for an $\cN=4$ $SU(N)$ SYM gas with temperature $T$
takes the form (\ref{pmgral}), with $R$ and $r_0$ the functions of
$N$ and $T$ specified in (\ref{Rr0}). We will now use this result
to determine the explicit form of the microscopic absorption
probabilities (\ref{pmsum}).

As we reviewed in Section \ref{reviewsec}, the model predicts that
the numbers of branes and antibranes are given by (\ref{nnbar}),
and the temperatures of the two gases are as indicated in
(\ref{tm}). Combining these with (\ref{Rr0}), we see that the
parameters to be used in (\ref{pmgral}) are
\be \label{dictionary}
R^2={1\over 2}r_h^2 e^{\alpha},\qquad r_0=2^{-1/4}r_h,\qquad
\Rb^2={1\over 2}r_h^2 e^{-\alpha},\qquad \rzb=2^{-1/4}r_{h}.
\ee
Plugging this into (\ref{pmgral}) and then (\ref{pmsum}), we get
our final prediction for the microscopic absorption probabilities,
\be \label{pm}
\Pm(\omega)={2^{-5l/2-15/4}\pi\over(l+2)!^{2}}\frac{\Gamma(1+l/4)^{4}}
{\Gamma(1+l/2)^{2}}(\omega r_h)^{2l+5}\cosh\alpha~.
\ee

\section{Comparison with Supergravity}
\label{sugrasec}

\subsection{Neutral case}
\label{neutralsubsec}

Let us now compare the microscopic predictions (\ref{pm}) against
the actual supergravity results, specializing first to the case of
the neutral black threebrane, $\alpha=0$, which is equivalent to the
Schwarzschild black hole in seven spacetime dimensions. The
corresponding absorption probabilities for arbitrary partial waves
of a minimal scalar field have been computed in 
\cite{kmr}:\footnote{The $l=0$ case had been worked out previously 
in \cite{dgm,emparan}.}
\be \label{pg0}
\Pg(\omega)={2^{-3l-3}\pi^2\over(l+2)!^{2}}\frac{\Gamma(1+l/4)^{2}}
{\Gamma(1/2+l/4)^{2}}(\omega r_h)^{2l+5}.
\ee
The functional dependence is in perfect agreement with (\ref{pm})
for $\alpha=0$. Despite appearances, using the identity
$\Gamma(x)=2^{1-2x}\sqrt{\pi}\,\Gamma(2x)/\Gamma(x+1/2)$ the
numerical coefficients can also be seen to agree, except for a
power of two:
\be \label{pcomp0}
\Pg(\omega)=2^{3/4+l/2}\Pm(\omega).
\ee
Notice that this numerical discrepancy can be summarized in
exactly the same manner as the one found for the entropy in
\cite{dgk}: if each gas could somehow carry twice the energy that
is available to it, then according to (\ref{em}) $T$ and $\Tb$
would increase\footnote{It is perhaps worth pointing out that even
in this case the Hawking and microscopic temperatures would not
agree, but instead $\Tm=2T_H$.} by a factor of $2^{1/4}$, which as
seen in (\ref{Rr0}) increases $r_0$ and $\rzb$ by the same factor,
implying in turn, through (\ref{pmgral}) and (\ref{pmsum}), that
$\Pm\to 2^{3/4+l/2}\Pm$. For $l=0$ this is not a new result, for
in fact the comparison in (\ref{pcomp0}) is, for the special case
of the s-wave, precisely the entropy comparison made in
\cite{dgk}: the absorption probabilities are of course
proportional to the corresponding cross sections, and for the
s-wave, the latter reduce at low frequencies to the respective
horizon areas \cite{dgm,emparan}, which are in turn, according to the
Bekenstein-Hawking formula, proportional to the entropies.

\subsection{Charged case}
\label{chargedsubsec}

To the best of our knowledge, the absorption probability for the
black threebrane arbitrarily far from extremality has not yet been
computed, so we will need to work it out here,
restricting again to the low-frequency regime (\ref{lowfreq}).
Fortunately, this will just amount to a simple generalization of the
calculation in Section \ref{symsubsec}.

The radial equation of motion is again (\ref{rhoeq}), now with the
replacements $\rho_0\to\rho_h$ and $\cR^4\to\rho_h^4
\sinh^2\alpha$, where $\rho_h\equiv\omega r_h$. In the \emph{outer
region} $\rho\gg\rho_h$ we still have the Bessel solution
(\ref{besselsol}). The difference is that now, contrary to
($\ref{regime}$), we do not have a clear separation between
$\rho_h$ and $\rho_h\sqrt{\sinh\alpha}$, and so we must define the
\emph{inner region} simply as $\rho_h\le\rho\ll 1$, implying that
both terms in the numerator of the last term of (\ref{rhoeq2}) are
comparable. It is still true, however, that the last term  of
(\ref{rhoeq2}) is relevant only very close to the horizon ($x=1$).
We conclude then that, at the order we are interested in, the only
change in the calculation is the modification of the exponent in
(\ref{phiansatz}) to
$$
-{i\over 4}\sqrt{{\rho_h^4\sinh^2\alpha\over\rho_h^2}+\rho_h^2}
=-{i\over 4}\rho_h\cosh\alpha~.
$$
Carrying this change through in (\ref{horflux}) and (\ref{pabs}),
and comparing with the result (\ref{pg0}) for the neutral case, we
deduce that for $\alpha\neq 0$ the absorption probability is
modified into
\be \label{pg}
\Pg(\omega)={2^{-3l-3}\pi^2\over(l+2)!^{2}}\frac{\Gamma(1+l/4)^{2}}
{\Gamma(1/2+l/4)^{2}}(\omega r_h)^{2l+5}\cosh\alpha~,
\ee
exactly as predicted by the microscopic result (\ref{pm})~!

We conclude then that, for arbitrary charge,
\be \label{pcomp}
\Pg(\omega)=2^{3/4+l/2}\Pm(\omega).
\ee
As explained in the previous subsection, this comparison was bound
to work for $l=0$, since in that case it is simply a rephrasing of
the entropy comparison in \cite{dgk}. The non-trivial results
obtained in this paper are the infinite number of successful
comparisons (\ref{pcomp}) for $l>0$.

\section{Hawking Radiation}
\label{emissionsec}

Given the absorption probabilities (\ref{pg}),
the corresponding absorption cross-sections follow as \cite{gubser}
\be \label{cp}
\cg(\omega)={8\pi^{2}\over
3\omega^{5}}(l+1)(l+2)^{2}(l+3)\Pg(\omega)~.
\ee
These in turn allow us to compute the rates of Hawking radiation into
each of the partial waves,
\be \label{Gg}
d\Gg(\omega)=\frac{\cg(\omega)\omega}{e^{\omega/T_{H}}-1}\,d\omega,
\ee
where $\cg(\omega)$ plays the role of greybody factor.

{}From the microscopic perspective, given that the brane and antibrane
subsystems are decoupled, we expect, in analogy to (\ref{pmsum}),
$$
d\Gm(\omega)=d\Gamma^{(l)}(\omega;N,T)+d\Gamma^{(l)}(\omega;\Nb,\Tb)~,
$$
where the rates on the right-hand side refer to bulk radiation
emerging from the two $\cN=4$ SYM gases. But again, through the
AdS/CFT correspondence, these should be equivalent to the rates of
emission for the corresponding \emph{near-extremal} black
threebranes, which are given by formulas analogous to (\ref{Gg}).
We thus have
\be \label{Gm}
d\Gm(\omega)=\left[\frac{\cm(\omega;N,T)}{e^{\omega/T}-1}
+\frac{\cm(\omega;\Nb,\Tb)}{e^{\omega/\Tb}-1}\right]\omega\,d\omega~,
\ee
which does not resemble (\ref{Gg}) in any obvious way.

In the low-frequency regime (\ref{lowfreq}) where we are working, the
supergravity and microscopic emission rates simplify to
\be \label{Gglow}
d\Gg(\omega)=\cg(\omega)T_{H}\,d\omega
\ee
and
\be \label{Gmlow}
d\Gm(\omega)=\left[\cm(\omega;N,T)T+\cm(\omega;N,T)\Tb\right]d\omega~,
\ee
respectively. Using (\ref{cp}), we see that
the comparison between these rates is
equivalent to the comparison of
$$
\Pg(\omega)T_{H}\quad \mbox{vs.}\quad
P^{(l)}(\omega;N,T)T+P^{(l)}(\omega;\Nb,\Tb)\Tb~,
$$
which is clearly independent from the successful match between
(\ref{pmsum}) and (\ref{pg}). Nevertheless, combining (\ref{pg})
and (\ref{tg}) we see that
$$
\Pg(\omega)T_{H}={2^{-3l-3}\pi\over(l+2)!^{2}}\frac{\Gamma(1+l/4)^{2}}
{\Gamma(1/2+l/4)^{2}}\omega^{2l+5} r_h^{2l+4}~,
$$
whereas using (\ref{pmgral}) with (\ref{Rr0}), (\ref{nnbar})
and (\ref{tm}) we obtain
$$
P^{(l)}(\omega;N,T)T+P^{(l)}(\omega;\Nb,\Tb)\Tb=
{2^{-5l/2-3}\over(l+2)!^{2}}\frac{\Gamma(1+l/4)^{4}}
{\Gamma(1+l/2)^{2}}\omega^{2l+5} r_h^{2l+4}~,
$$
so that we again have a perfect functional match!

Just like in the previous section, despite their superficial
dissimilarity the numerical coefficients also agree, except for
the power of two that corresponds to doubling the energy of each
gas (which translates into increasing $T$ and $\Tb$ by a factor of
$2^{1/4}$, and setting $r_{0}=\rzb=r_{h}$, rather than
$2^{-1/4}r_{h}$ as in (\ref{dictionary})). In short, we have found
that
\be \label{Gcomp}
\Gg(\omega)=2^{l/2}\Gm(\omega).
\ee

Equally important, we have learned that, at least to lowest order
in the frequency, the separate Hawking radiation rates for the
D3-brane and \aD-brane stacks \emph{agree} with one another,
\be \label{Gddbar}
d\Gamma^{(l)}(\omega;N,T)=d\Gamma^{(l)}(\omega;\Nb,\Tb)~.
\ee
This is in spite of the fact that the two gases have
\emph{different} temperatures. As a matter of fact, using
(\ref{Rr0}) in (\ref{pmgral}) one sees that the product
$P^{(l)}(\omega;N,T)T$ which controls the D3-brane emission rate
depends on $N$ and $T$ only through the combination $N^{2}T^{4}$,
which is of course the energy density (\ref{em}) of the
corresponding gas. We conclude then that the D3 and \aD\ radiation
rates agree precisely because the two gases have \emph{equal}
energies! At the very least, this is an important self-consistency
check for the model: since one postulates that the two gas
energies are the same, it is satisfying to see that this equality
will not be disturbed when the black brane radiates, which is part
of what it does for a living. But one can actually view this as an
\emph{explanation} of the equal-energy condition: if the energies
were initially different (for instance, on account of the gases
having equal temperatures), then the gas with higher energy would
radiate more, and the energies would tend to equalize. It is only
the equal-energy case that is in this sense `stable'.

On the other hand, we have seen in the previous section that the
D3 and \aD\ absorption probabilities are in fact \emph{different},
which implies that, when we disturb the black brane by throwing
some radiation at it, the component with the lowest temperature
(or, equivalently, the largest number of branes) will absorb more
energy. This suggests that the supergravity counterpart of the
microscopic system with unequal gas energies is some type of
excited state of the black brane, which will eventually relax back
the preferred equal-energy configuration. Notice that the
corresponding solution will in general not possess the same
symmetry properties as the original threebrane solution. For
instance, after absorption of $l>0$ radiation one would expect the
black brane to become distorted into some configuration that is no
longer spherically symmetric.

It would clearly be of great interest to establish whether the
above findings continue to hold at next-to-leading (or perhaps
even higher) order in the radiation frequency, or if they are just
somehow a special property of the lowest-order terms. Leaving a
detailed study of this question for future work, let us just
remark at this point that at higher order the stability analysis
would be more involved, for one would for instance have to take
into account the possibility that part of the radiation emitted by
one of the stacks is absorbed by the other.

\section{Conclusions}
\label{conclsec}

\subsection{Summary of results}
\label{summarysubsec}

We have demonstrated that the brane-antibrane model formulated in
\cite{dgk} can correctly account for the low-frequency absorption
probabilities and Hawking emission rates of the black threebrane
arbitrarily far from extremality, for arbitrary partial waves of a
minimal scalar field. Our main results, the comparisons
(\ref{pcomp}) and (\ref{Gcomp}), amount to an infinite number of
new tests of the microscopic model. Notice that these tests are
indeed independent from one another. One might for instance
suspect that since the passage from one partial wave to the next
involves an additional derivative in the
brane-bulk coupling, it is bound to give rise to
a factor of $\omega^{2}$, and
by dimensional analysis, $r_h^{2}$, in the absorption
probability. In this way all of the absorption results for higher
partial waves would be related to the $l=0$ case, which, as
explained in Section \ref{neutralsubsec}, is in fact nothing but
the entropy comparison in \cite{dgk}. That this is in general
not the whole
story can be seen by noting that the $l$-dependence of the
exponent of the frequency is \emph{not} the same in, for instance,
the extremal ($P\propto\omega^{4l+8}$) \cite{klebabs,partiald3abs} 
and
near-extremal ($P\propto\omega^{2l+5}$) \cite{ps} cases. And in
any event, dimensional analysis obviously does not control the
comparison between the numerical coefficients in the microscopic
and supergravity sides, which has been seen to be successful for
all partial waves, up to the same factor of 2 in the gas energies.

It is interesting that, as can be seen by combining (\ref{em}) and
(\ref{Rr0}), the equal-energy condition amounts to the statement
that the horizon radii of the two near-extremal branes employed in
the microscopic side coincide, $r_0=\rzb$, and then the rescaling
needed to resolve the numerical discrepancy identifies these with
the horizon radius of the brane in the supergravity side,
$r_0=\rzb=r_h$, as noted for instance above (\ref{Gcomp}). This is
essentially the agreement \cite{dgk,sp,bl} mentioned in the
Introduction between $r_h$ and the transverse size of the
microscopic system, $\sqrt{\expec{\Phi^2}}\,$, except that here we
\emph{are} keeping track the numerical coefficient.

One possible source of confusion is the fact that, as in previous
analyses of the model, we have employed supergravity for the
calculations in the \emph{microscopic} side (see Sections
\ref{symsubsec} and \ref{emissionsec}). More concretely, what
could seem suspicious in this paper is that the microscopic and
supergravity absorption probabilities are obtained by solving the
\emph{same} equation of motion, namely, (\ref{rhoeq}). One should
not however lose sight of the fact that, as was emphasized in
\cite{dgk,ghm}, the results to be compared are extracted from two
completely different regimes (near-extremal vs.
arbitrarily-far-from-extremal) of (\ref{rhoeq}). Moreover, in the
microscopic side the parameters $r_0$ and $R$ are not chosen in an
\emph{ad hoc} manner, but fixed by a maximization procedure. (The
only condition that \emph{is} imposed by hand is the equality of
the energies of the two gases.) And, perhaps most significant of
all, we do not simply compare one supergravity absorption
probability against another, but one against the \emph{sum} of two
others, in a setup where the three corresponding brane charges are
in general all \emph{different} from one another.

To bring out more clearly the precise sense in which the agreement
found in the absorption calculation is non-trivial, imagine we had
access to the \emph{exact} absorption probability that follows
from the equation of motion (\ref{rhoeq}), which we could denote
as $P_{(l)}(\omega;r_0,(R/r_{0})^2)$. Then the supergravity
probability is of course
$$
\Pg(\omega)=P_{(l)}(\omega;r_h,\sinh\alpha)~.
$$
The prescription of the model, on the other hand, proceeds in
three steps. We first consider the near-extremal limit of the full
absorption probability,
$P^{\mathrm{NE}}_{(l)}(\omega;r_0,(R/r_{0})^2)$, which by
definition is \emph{just the leading term} of
$P_{(l)}(\omega;r_0,(R/r_{0})^2)$ for arbitrarily large (but
finite) $(R/r_{0})^{2}$. Second, we apply this formula separately
to the brane and antibrane subsystems with the parameters
predicted by the model, namely, (\ref{dictionary}). To avoid the
numerical discrepancy, for the purpose of this discussion we will
from the start set $r_0=\rzb=r_h$, rather than $2^{-1/4}r_h$.
Third, the microscopic absorption probability is obtained from the
sum of the brane and antibrane contributions, i.e.,
$$
\Pm(\omega)=P^{\mathrm{NE}}_{(l)}(\omega;r_h,e^{\alpha}/2)
+P^{\mathrm{NE}}_{(l)}(\omega;r_h,e^{-\alpha}/2)~.
$$
For $P_{(l)}(\omega;r_h,\sinh\alpha)$ an arbitrary function, there
is clearly no reason whatsoever for $\Pm(\omega)$ to agree with
$\Pg(\omega)$.  Nevertheless, we have found that, to lowest order
in the frequency, $P_{(l)}(\omega;r_0,(R/r_{0})^2)$ takes the form
$p(\omega r_{0})\sqrt{1+(R/r_{0})^4}$, which is precisely as
required to pass this test. It is far from obvious whether this
pattern can continue to hold at higher order.

As we have emphasized in Section \ref{emissionsec}, the comparison
of the Hawking radiation rates brings in an entirely new
requirement. {}From a physical perspective, the agreement
(\ref{Gcomp}) between the microscopic and supergravity emission
rates is perhaps even more striking than that between the
absorption probabilities, in particular because it is completely
independent from the entropy match found in \cite{dgk}. Notice,
however, that at least at this order the black brane does not
strictly speaking `know' that it is made of two independent
components that emit radiation independently, because the two
different temperatures are not \emph{directly} visible in the
supergravity side. This is unlike the situation in, e.g., the
D1-D5 system, where inclusion of the greybody factor is explicitly
seen to convert the single thermal factor for emission at the
Hawking temperature into the product (rather than the sum, as we
have in our case) of two thermal factors corresponding to
different temperatures \cite{ms}. The difference between the two
cases is of course due to the emission mechanism: while in
\cite{ms} the emission of a massless closed string into the bulk
necessarily involves an interaction between the two gases, in our
case the two gases are free to radiate independently. A
higher-order calculation might conceivably allow one to see the
two independent temperatures somewhat more explicitly in the
supergravity side, but of course it is far from clear whether the
agreement between the microscopic and supergravity emission rates
can persist in such a calculation.

{}From the Hawking radiation analysis we have also learned that
(at least for low frequencies) the  rate of emission for each of
the two components of the microscopic system depends only on the
energy of the corresponding gas. If the gas energies were not
equal, then the gas with higher energy would radiate more, and so
the energies would tend to equalize. Our results can therefore be
viewed as an explanation for the equal-energy condition, which up
to now has been simply a postulate of the model. As we have also
discussed in Section \ref{emissionsec}, however, the fact that the
D3-brane and \aD-brane absorption probabilities are different
implies that it should be possible to achieve unequal gas energies
when we disturb the system by throwing radiation at it. This in
turn suggests that the supergravity counterpart of the microscopic
configuration with different gas energies should be some type of
excited state of the black brane.

As has been noted already at various points in the above
discussion, it is an important outstanding problem to establish
whether the absorption and emission probabilities continue to
agree at higher order in the radiation frequency. We hope to
report on this question in future work.

To summarize, in this paper we have found significant new evidence
in favor of the brane-antibrane model, and have moreover thrown
light on the equal-energy condition which was heretofore one of
its most puzzling aspects.

\subsection{Critical assessment of the model}
\label{criticalsubsec}

In spite of the successes of the model (which include in
particular  the results reported in this paper), the situation is
not yet entirely satisfactory, and more work will be needed to
conclusively validate the model. In preparation for it, it seems
useful to include here a list of the various assumptions of the
model that could be in need of more careful scrutiny:

\begin{enumerate}
\item The model assumes that the D3-\aD\ pairs do not annihilate
even partially, or in other words, that the open string vacuum is,
at the relevant temperatures, stable. A calculation supporting
this assumption was included in \cite{dgk}, and additional
evidence has been provided by the numerical analysis in \cite{sp}.
As mentioned in the Introduction, the interpretation of
\cite{kalyanarama,ks} can perhaps be viewed as a relaxation of
this assumption.

\item It is assumed that all $3$-$\overline{3}$ open strings
acquire large masses and consequently cannot be excited. In other
words, the model incorporates gases with $\cO(N^2)$ and
$\cO(\Nb^2)$ degrees of freedom, but not $\cO(N\Nb)$. The lowest
$3$-$\overline{3}$ mode is of course the tachyon, which must by
definition acquire a large positive mass-squared if the
brane-antibrane system is to be stable. One must also consider the
$3$-$\overline{3}$ fermions (arising from the Ramond sector) that
would at zero temperature be massless. The analysis of \cite{dgk}
provided some evidence that both of these modes indeed become
massive enough to decouple.

\item The model does not include any binding energies. As
emphasized in \cite{sp}, at least naively one would in fact expect
this to be wrong. In particular, it seems difficult to see how
(e.g., gravitational) D3-\aD\ binding energy could be avoided.
This would be expected to arise from closed string exchange
between the two stacks, or equivalently, from loops of
$3$-$\overline{3}$ open strings (which could contribute through
virtual effects even if on-shell they are postulated to have large
masses). If present, this type of binding energy would bring in
some $N\Nb$-dependence.

\item The model involves a restriction on the types of
brane-antibrane pairs that contribute: even arbitrarily far from
extremality it is assumed that there is no contribution from
D1-$\overline{\mbox{D1}}$ pairs. To understand this, it seems
natural to try to extend the model to the case of the black brane
which has \emph{both} five-form and three-form RR charge
\cite{d3d1sol}. A preliminary analysis appears to indicate that
indeed D1-$\overline{\mbox{D1}}$ pairs disappear altogether when
the three-form charge is taken to approach zero \cite{d3d1}. Other
potentially relevant observations may be found in
\cite{d3d1other}.

\item The model assumes that the component gases have equal
energies, and consequently different temperatures. It is perhaps
worth pointing out that one \emph{can} in fact reproduce the
supergravity entropy with a model based on equal-temperature
gases, as long as one is willing to pay the price of fixing the
number of branes and antibranes by hand, instead of choosing the
value of $N$ that maximizes the entropy. In this new scenario one
would also lose many of the other successful predictions of the
equal-energy model. It is therefore satisfying that the results of
the present paper appear to provide for the first time a rationale
for the equal-energy condition, in terms of `stability' of the
system with respect to Hawking emission.

\item The model assumes a specific form for the dynamics of the
relevant gases: the formulas employed are those of
strongly-coupled $\cN=4$ SYM (or equivalently, a near-extremal
black threebrane). Since the relevant temperatures are much lower
than the string scale one would naively expect $\ap$ corrections
to be suppressed. But in fact, when one writes down, e.g., the
Born-Infeld action, there are some additional factors of $N$
hidden in the non-Abelian nature of the field strength, which can
be seen to imply that the higher-derivative corrections are
controlled not by $\ls$ but by the scale $R=(4\pi\gs N)^{1/4}\ls$,
which is of course the throat radius of the corresponding
supergravity solution \cite{klebabs,ghkk,dealwis}. What is very
peculiar about the model is that, as can be seen by comparing
(\ref{tm}) against (\ref{dictionary}), the predicted gas
temperatures are precisely of order $1/R$, and so \emph{a priori}
one would have expected the higher-derivative corrections to play
an important role,\footnote{It has been argued in \cite{gh,i} (see
also \cite{otherd3holo,dgks,ejp}) that at strong-coupling only the
$R^4 F^{4}$ correction is present. However, that argument relies
heavily on supersymmetry, which is of course broken in our finite
temperature setup. The effect of the temperature has been studied
in \cite{ejp}.} in which case we would not be entitled to employ
the SYM/near-extremal formulas.

\end{enumerate}

It is clear from this list that the status of the brane-antibrane
model is still open to debate. At the same time, the list
certainly makes the body of evidence that has by now accumulated
in favor of the model seem all the more remarkable.

The two most questionable assumptions are items 3 and 6. Notice
that these two points are in fact not unrelated: the
higher-derivative corrections of item 6 arise from massive
open string modes, whose cummulative effect in
loops is equivalent to closed string exchange,
and is therefore associated with binding energy, this time of D3-D3 or
\aD-\aD\ type. A remarkable fact is that, if one attempts to
incorporate the effect of these expected higher-derivative
corrections into the energy and entropy formulas by dimensional
analysis,\footnote{One might again have a chance of extracting the
correct formulas for the strongly-coupled gas from supergravity,
by invoking the generalization of AdS/CFT to higher energies, as
in \cite{klebabs},\cite{ghkk}-\cite{acgg}. Notice, however, that
one should \emph{not} simply use the formulas for the
(non-near-extremal) black threebrane, since according to the model
(and physical intuition) that would include not only the desired
contribution from the gas, but also from additional D3-\aD\ pairs
(which is precisely why one gets, for instance, negative specific
heat).} then \emph{after} the maximization procedure one concludes
that they modify the final entropy-mass relation only through a
numerical factor!\footnote{We thank Mart\'{\i}n Kruczenski for
emphasizing this point to us.} This might very well be, then, the
origin of the ubiquitous factor of 2.

\section*{Acknowledgements}

We are grateful to Elena C\'aceres, Alejandro Corichi, Ulf
Danielsson, Mart\'{\i}n Kruczenski, and David Lowe for useful
discussions. This work was partially supported by Mexico's
National Council of Science and Technology (CONACyT), under grant
CONACyT-40745-F, and by DGAPA-UNAM, under grant IN104503-3.


\begin{thebibliography}{99}


\bibitem{dgk}
U. Danielsson, A. G\"uijosa, M. Kruczenski, ``Brane-Antibrane
systems at finite temperature and the
 entropy of black holes,''
 JHEP {\bf 0109} (2001) 011, [arXiv:hep-th/0106201].

\bibitem{dgk2}
U.~H.~Danielsson, A.~G\"uijosa and M.~Kruczenski, ``Black brane
entropy from brane-antibrane systems,'' Rev.\ Mex.\ F\'{\i}s.\
{\bf 49S2} (2003) 61 [arXiv:gr-qc/0204010].

\bibitem{ghm}
A.~G\"uijosa, H.~H.~Hern\'andez Hern\'andez and H.~A.~Morales
T\'ecotl, ``The entropy of the rotating charged black threebrane
from a brane-antibrane system,'' JHEP {\bf 0403} (2004) 069
[arXiv:hep-th/0402158].

\bibitem{sp}
O.~Saremi and A.~W.~Peet, ``Brane-antibrane systems and the
thermal life of neutral black holes,'' arXiv:hep-th/0403170.

\bibitem{bl}
O.~Bergman and G.~Lifschytz, ``Schwarzschild black branes from
unstable D-branes,'' JHEP {\bf 0404}, 060 (2004)
[arXiv:hep-th/0403189].

\bibitem{kalyanarama}
S.~Kalyana Rama, ``A description of Schwarzschild black holes in
terms of intersecting M-branes and antibranes,''
arXiv:hep-th/0404026.

\bibitem{lifschytz}
G.~Lifschytz, ``Charged black holes from near extremal black
holes,'' arXiv:hep-th/0405042.

\bibitem{ks}
S.~K.~Rama and S.~Siwach, ``A description of multicharged black
holes in terms of branes and antibranes,'' arXiv:hep-th/0405084.

\bibitem{halyo}
E.~Halyo, ``Black hole entropy and superconformal field theories
on brane-antibrane systems,'' arXiv:hep-th/0406082.

\bibitem{hms}
G.~T.~Horowitz, J.~M.~Maldacena and A.~Strominger, ``Nonextremal
Black Hole Microstates and U-duality,'' Phys.\ Lett.\ B {\bf 383}
(1996) 151, [arXiv:hep-th/hep-th/9603109].

\bibitem{antientropy}
M.~Cveti\v{c} and D.~Youm, ``Entropy of Non-Extreme Charged
Rotating Black Holes in String Theory,'' Phys.\ Rev.\ D {\bf 54}
(1996)
2612, [arXiv:hep-th/9603147];\\
G.~T.~Horowitz, D.~A.~Lowe and J.~M.~Maldacena, ``Statistical
Entropy of Nonextremal Four-Dimensional Black Holes and
U-Duality,'' Phys.\ Rev.\ Lett.\  {\bf 77} (1996) 430,
[arXiv:hep-th/9603195]; \\
R.~Kallosh and A.~Rajaraman, ``Brane-anti-brane Democracy,''
Phys.\ Rev.\ D {\bf 54} (1996) 6381,
[arXiv:hep-th/9604193]; \\
J.~Zhou, H.~J.~M\"uller-Kirsten, J.~Q.~Liang and F.~Zimmerschied,
``M-branes, anti-M-branes and non-extremal black holes,'' Nucl.\
Phys.\ B {\bf 487} (1997) 155
[arXiv:hep-th/9611146]; \\
M.~S.~Costa and M.~J.~Perry, ``Landau degeneracy and black hole
entropy,'' Nucl.\ Phys.\ B {\bf 520} (1998) 205,
[arXiv:hep-th/9712026].

\bibitem{kll}
D.~Kabat, G.~Lifschytz and D.~A.~Lowe,
 ``Black hole thermodynamics from calculations in strongly coupled gauge
theory,'' Phys.\ Rev.\ Lett.\  {\bf 86} (2001) 1426 [Int.\ J.\
Mod.\ Phys.\
A {\bf 16} (2001) 856] [arXiv:hep-th/0007051];\\
D.~Kabat, G.~Lifschytz and D.~A.~Lowe, ``Black hole entropy from
non-perturbative gauge theory,'' Phys.\ Rev.\ D {\bf 64} (2001)
124015 [arXiv:hep-th/0105171];\\
N.~Iizuka, D.~Kabat, G.~Lifschytz and D.~A.~Lowe, ``Probing black
holes in non-perturbative gauge theory,'' Phys.\ Rev.\ D {\bf 65}
(2002) 024012 [arXiv:hep-th/0108006].

\bibitem{mathur}
O.~Lunin and S.~D.~Mathur, ``Statistical interpretation of
Bekenstein entropy for systems with a stretched horizon,'' Phys.\
Rev.\ Lett.\  {\bf 88} (2002) 211303 [arXiv:hep-th/0202072];\\
S.~D.~Mathur, A.~Saxena and Y.~K.~Srivastava, ``Constructing
'hair' for the three charge hole,''
Nucl.\ Phys.\ B {\bf 680} (2004) 415 [arXiv:hep-th/0311092];\\
S.~D.~Mathur, ``Where are the states of a black hole?,''
arXiv:hep-th/0401115.


\bibitem{glthermo} S.~S.~Gubser and I.~Mitra, ``Instability of
charged black holes in anti-de Sitter space,''
{\tt hep-th/0009126}; \\
S.~S.~Gubser and I.~Mitra, `The evolution of unstable black holes
in anti-de Sitter space,''
{\tt hep-th/0011127}; \\
H.~S.~Reall, ``Classical and thermodynamic stability of black
branes,'' {\tt hep-th/0104071};\\
J.~P.~Gregory and S.~F.~Ross,
``Stability and the negative mode for Schwarzschild in a finite cavity,''
Phys.\ Rev.\ D {\bf 64} (2001) 124006
[arXiv:hep-th/0106220].

\bibitem{gl}
R.~Gregory and R.~Laflamme, ``Black strings and p-branes are
unstable,'' Phys.\ Rev.\ Lett.\  {\bf 70} (1993) 2837
{\tt hep-th/9301052}; \\
R.~Gregory and R.~Laflamme, ``The Instability of charged black
strings and p-branes,'' Nucl.\ Phys.\ B {\bf 428} (1994) 399, {\tt
hep-th/9404071}.


\bibitem{lifschytz2}
G.~Lifschytz, ``Black hole thermalization rate from brane
anti-brane model,'' arXiv:hep-th/0406203.

\bibitem{ikll}
N.~Iizuka, D.~Kabat, G.~Lifschytz and D.~A.~Lowe, ``Quasiparticle
picture of black holes and the entropy-area relation,'' Phys.\
Rev.\ D {\bf 67} (2003) 124001 [arXiv:hep-th/0212246];\\
N.~Iizuka, D.~Kabat, G.~Lifschytz and D.~A.~Lowe, ``Stretched
horizons, quasiparticles and quasinormal modes,'' Phys.\ Rev.\ D
{\bf 68} (2003) 084021 [arXiv:hep-th/0306209].

\bibitem{halyobits}
E.~Halyo, ``Gravitational entropy and string bits on the stretched
horizon,'' arXiv:hep-th/0308166.

\bibitem{gkp}
S.~S.~Gubser, I.~R.~Klebanov and A.~W.~Peet, ``Entropy and
Temperature of Black 3-Branes,'' Phys.\ Rev.\ D {\bf 54} (1996)
3915, [arXiv:hep-th/9602135].

\bibitem{threequarters}
S.~S.~Gubser, I.~R.~Klebanov and A.~M.~Polyakov, ``Gauge theory
correlators from non-critical string theory,'' Phys.\ Lett.\ B
{\bf 428} (1998) 105,
[arXiv:hep-th/9802109]; \\
S.~S.~Gubser, I.~R.~Klebanov and A.~A.~Tseytlin, ``Coupling
constant dependence in the thermodynamics of $\cN = 4$
supersymmetric Yang-Mills theory,'' Nucl.\ Phys.\ B {\bf 534}
(1998) 202, [arXiv:hep-th/9805156].

\bibitem{malda}
J.~Maldacena, ``The large $N$ limit of superconformal field
theories and supergravity,'' Adv.\ Theor.\ Math.\ Phys.\  {\bf 2},
231 (1998) [Int.\ J.\ Theor.\ Phys.\  {\bf 38}, 1113 (1998)],
[arXiv:hep-th/9711200].

\bibitem{ps}
G.~Policastro and A.~Starinets, ``On the absorption by
near-extremal black branes,'' Nucl.\ Phys.\ B {\bf 610} (2001) 117
[arXiv:hep-th/0104065].

\bibitem{d3absorption}
Y.~Satoh,
 ``Propagation of scalars in non-extremal black hole and black p-brane
geometries,'' Phys.\ Rev.\ D {\bf 58}, 044004 (1998)
[arXiv:hep-th/9801125];\\
S.~Musiri and G.~Siopsis, ``Temperature of D3-branes off
extremality,'' Phys.\ Lett.\ B {\bf 504}, 314 (2001)
[arXiv:hep-th/0003284];\\
J.~F.~V\'azquez-Poritz, ``Absorption by nonextremal D3-branes,''
arXiv:hep-th/0007202.

\bibitem{gr}
I.~S.~Gradshteyn and I.~M.~Ryzhik, {\it Tables of Integrals,
Series, and Products}, Academic Press (1994).

\bibitem{kmr}
P.~Kanti and J.~March-Russell, ``Calculable corrections to brane
black hole decay. I: The scalar case,'' Phys.\ Rev.\ D {\bf 66}
(2002) 024023 [arXiv:hep-ph/0203223].

\bibitem{dgm}
S.~R.~Das, G.~W.~Gibbons and S.~D.~Mathur, ``Universality of low
energy absorption cross sections for black holes,'' Phys.\ Rev.\
Lett.\  {\bf 78}, 417 (1997) [arXiv:hep-th/9609052].

\bibitem{emparan}
R.~Empar\'an,
``Absorption of scalars by extended objects,''
Nucl.\ Phys.\ B {\bf 516} (1998) 297
[arXiv:hep-th/9706204].

\bibitem{gubser}
S.~S.~Gubser, ``Can the effective string see higher partial waves?,''
Phys.\ Rev.\ D {\bf 56}, 4984 (1997) [arXiv:hep-th/9704195].

\bibitem{klebabs}
I.~R.~Klebanov, ``World-volume approach to absorption by
non-dilatonic branes,'' Nucl.\ Phys.\ B {\bf 496} (1997) 231
[arXiv:hep-th/9702076].

\bibitem{partiald3abs}
S.~S.~Gubser, I.~R.~Klebanov and A.~A.~Tseytlin,
``String theory and classical absorption by three-branes,''
Nucl.\ Phys.\ B {\bf 499}, 217 (1997)
[arXiv:hep-th/9703040];\\
I.~R.~Klebanov, W.~I.~Taylor and M.~Van Raamsdonk,
``Absorption of dilaton partial waves by D3-branes,''
Nucl.\ Phys.\ B {\bf 560}, 207 (1999)
[arXiv:hep-th/9905174].

\bibitem{ms}
J.~M.~Maldacena and A.~Strominger, ``Black hole greybody factors
and D-brane spectroscopy,'' Phys.\ Rev.\ D {\bf 55} (1997) 861
[arXiv:hep-th/9609026].

\bibitem{d3d1sol}
J.~G.~Russo and A.~A.~Tseytlin, ``Waves, boosted branes and BPS
states in M-theory,'' Nucl.\ Phys.\ B {\bf 490} (1997) 121
[arXiv:hep-th/9611047];\\
J.~X.~Lu and S.~Roy, ``((F, D1), D3) bound state and its T dual
daughters,'' JHEP {\bf 0001}, 034 (2000) [arXiv:hep-th/9905014].

\bibitem{d3d1}
A.~G\"uijosa, unpublished.

\bibitem{d3d1other}
J.~M.~Maldacena and J.~G.~Russo,
``Large N limit of non-commutative gauge theories,''
JHEP {\bf 9909} (1999) 025
[arXiv:hep-th/9908134];\\
R.~G.~Cai and N.~Ohta, ``On the thermodynamics of large N
non-commutative super Yang-Mills  theory,'' Phys.\ Rev.\ D {\bf
61}, 124012 (2000) [arXiv:hep-th/9910092];\\
``Noncommutative and ordinary super Yang-Mills on (D(p-2),Dp) bound  states,''
JHEP {\bf 0003}, 009 (2000)
[arXiv:hep-th/0001213];\\
D.~Youm, ``A note on (D(p-2),Dp) bound state and noncommutative
Yang-Mills theory,'' Mod.\ Phys.\ Lett.\ A {\bf 15}, 1949 (2000)
[arXiv:hep-th/0006019].

\bibitem{ghkk}
S.~S.~Gubser, A.~Hashimoto, I.~R.~Klebanov and M.~Krasnitz,
``Scalar absorption and the breaking of the world volume conformal
invariance,'' Nucl.\ Phys.\ B {\bf 526} (1998) 393
[arXiv:hep-th/9803023].

\bibitem{dealwis}
S.~P.~de Alwis, ``Supergravity, the DBI action and black hole
physics,'' Phys.\ Lett.\ B {\bf 435} (1998) 31
[arXiv:hep-th/9804019].

\bibitem{gh}
S.~S.~Gubser and A.~Hashimoto, ``Exact absorption probabilities
for the D3-brane,'' Commun.\ Math.\ Phys.\  {\bf 203} (1999) 325
[arXiv:hep-th/9805140].

\bibitem{i}
K.~A.~Intriligator, ``Maximally supersymmetric RG flows and AdS
duality,'' Nucl.\ Phys.\ B {\bf 580} (2000) 99
[arXiv:hep-th/9909082].

\bibitem{otherd3holo}
N.~R.~Constable and R.~C.~Myers, ``Exotic scalar states in the
AdS/CFT correspondence,'' JHEP {\bf 9911} (1999) 020
[arXiv:hep-th/9905081];\\
M.~S.~Costa, ``Absorption by double-centered D3-branes and the
Coulomb branch of N = 4  SYM theory,'' JHEP {\bf 0005} (2000) 041
[arXiv:hep-th/9912073]; \\
M.~S.~Costa, ``A test of the AdS/CFT duality on the Coulomb
branch,'' Phys.\ Lett.\ B {\bf 482} (2000) 287 [Erratum-ibid.\ B
{\bf 489} (2000) 439] [arXiv:hep-th/0003289];\\
L.~Rastelli and M.~Van Raamsdonk, ``A note on dilaton absorption
and near-infrared D3 brane holography,'' JHEP {\bf 0012}, 005
(2000) [arXiv:hep-th/0011044].

\bibitem{dgks}
U.~H.~Danielsson, A.~G\"uijosa, M.~Kruczenski and B.~Sundborg,
``D3-brane holography,'' JHEP {\bf 0005} (2000) 028
[arXiv:hep-th/0004187].

\bibitem{ejp}
N.~J.~Evans, C.~V.~Johnson and M.~Petrini, ``Clearing the throat:
Irrelevant operators and finite temperature in  large N gauge
theory,'' JHEP {\bf 0205} (2002) 002 [arXiv:hep-th/0112058].

\bibitem{acgg}
X.~Amador, E.~C\'aceres, H.~Garc\'{\i}a-Compe\'an and
A.~G\"uijosa, ``Conifold holography,'' JHEP {\bf 0306} (2003) 049
[arXiv:hep-th/0305257].

\end{thebibliography}
\end{document}